\documentclass[11pt,a4paper]{article}
\usepackage[hyperref]{emnlp2020}
\usepackage{times}
\usepackage{latexsym}

\usepackage{graphicx}
\usepackage{subcaption}
\usepackage{amsmath}
\usepackage{pgfplots}
\usepackage{booktabs}
\usepackage[T1]{fontenc}

\usepackage{microtype}

\aclfinalcopy 
\def\aclpaperid{2218} 

\title{Digital Voicing of Silent Speech}

\author{David Gaddy \and Dan Klein \\
  University of California, Berkeley \\
  \texttt{\{dgaddy,klein\}@berkeley.edu}
   \\}

\date{}

\definecolor{cornflowerblue}{HTML}{4285F4}
\newcommand{\norm}[1]{\left\lVert#1\right\rVert}

\begin{document}
\maketitle
\begin{abstract}
In this paper, we consider the task of digitally voicing silent speech, where silently mouthed words are converted to audible speech based on electromyography (EMG) sensor measurements that capture muscle impulses.
While prior work has focused on training speech synthesis models from EMG collected during \emph{vocalized} speech, we are the first to train from EMG collected during silently articulated speech.
We introduce a method of training on silent EMG by transferring audio targets from vocalized to silent signals.
Our method greatly improves intelligibility of audio generated from silent EMG compared to a baseline that only trains with vocalized data, decreasing transcription word error rate from 64\% to 4\% in one data condition and 88\% to 68\% in another.
To spur further development on this task, we share our new dataset of silent and vocalized facial EMG measurements.
\end{abstract}

\section{Introduction}

In this paper, we are interested in in enabling speech-like communication without requiring sound to be produced.
By using muscular sensor measurements of speech articulator movement, we aim to capture silent speech - utterances that have been articulated without producing sound.
In particular, we focus on the task which we call \emph{digital voicing}, or generating synthetic speech to be transmitted or played back.

Digitally voicing silent speech has a wide array of potential applications.  For example, it could be used to create a device analogous to a Bluetooth headset that allows people to carry on phone conversations without disrupting those around them.  Such a device could also be useful in  settings where the environment is too loud to capture audible speech or where maintaining silence is important.  Alternatively, the technology could be used by some people who are no longer able to produce audible speech, such as individuals whose larynx has been removed due to trauma or disease \cite{meltzner2017silent}.
In addition to these direct uses of digital voicing for silent speech, it may also be useful as a component technology for creating silent speech-to-text systems \cite{schultz2010modeling}, making silent speech accessible to our devices and digital assistants by leveraging existing high-quality audio-based speech-to-text systems.

\begin{figure}
    \centering
    \scalebox{-1}[1]{\includegraphics[width=.8\columnwidth]{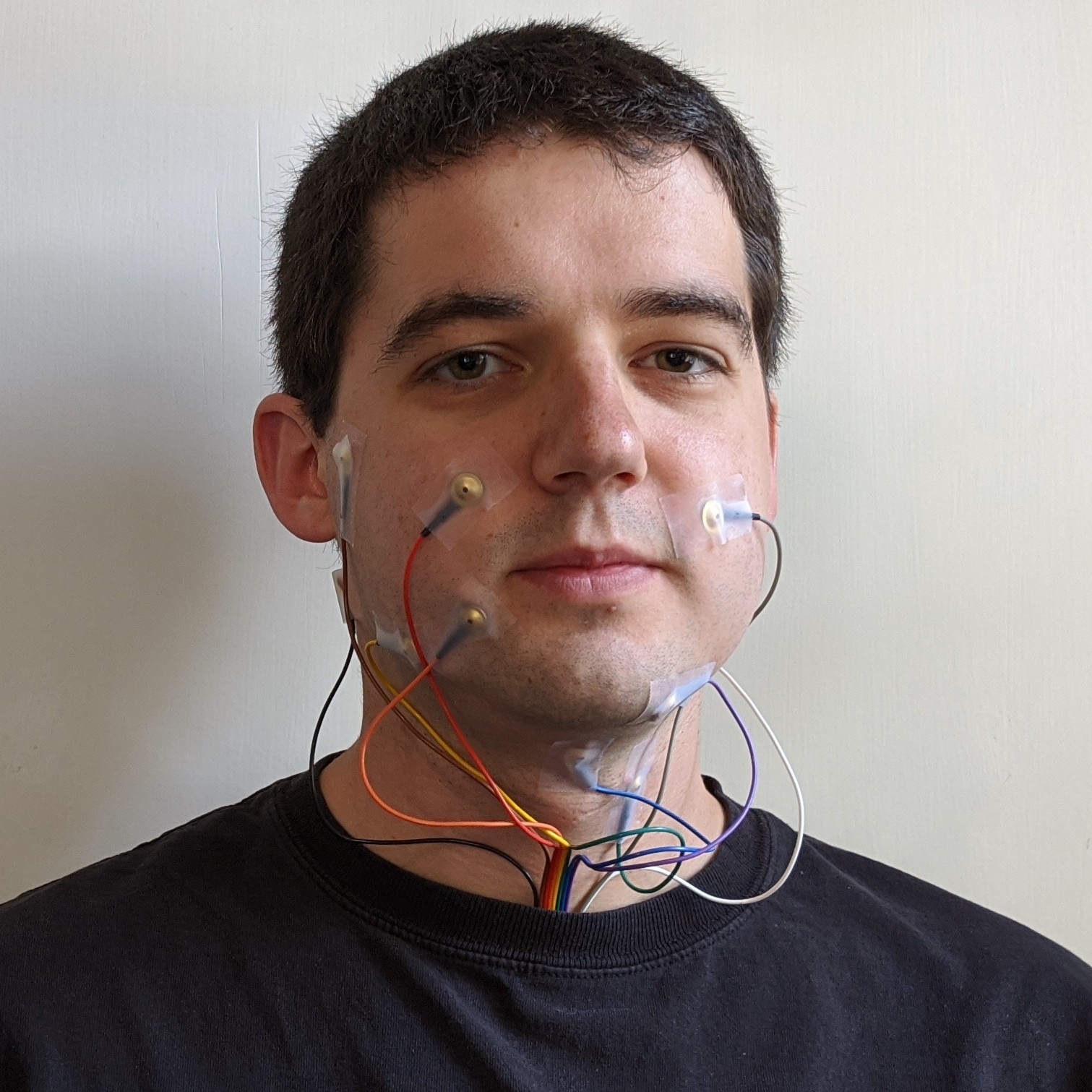}}
    \caption{Electromyography (EMG) electrodes placed on the face can detect muscle movements from speech articulators.}
    \label{fig:emg-pic}
\end{figure}

\begin{figure*}
    \captionsetup[subfigure]{labelformat=empty}
    \begin{subfigure}[b]{0.5\textwidth}
    \includegraphics[width=0.9\columnwidth]{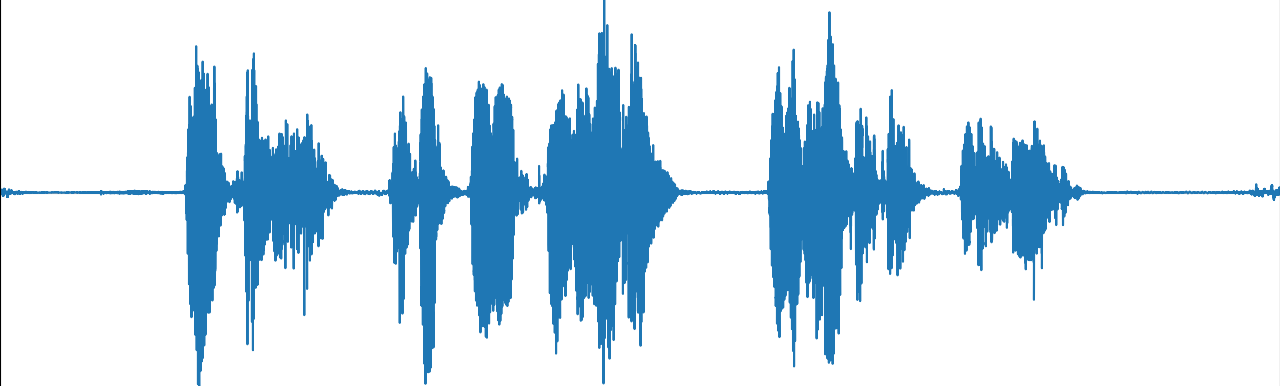}
    \caption{{\large $A_V$ } - audio from vocalized speech}
    \end{subfigure}
    \begin{subfigure}[b]{0.5\textwidth}
    \hspace{4em}
    \includegraphics[width=0.3\columnwidth]{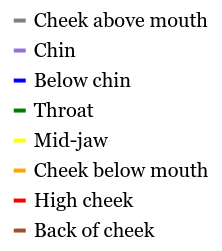}
    \vspace{.5em}
    \end{subfigure}\\
    \begin{subfigure}[b]{0.5\textwidth}
    \includegraphics[width=0.9\columnwidth]{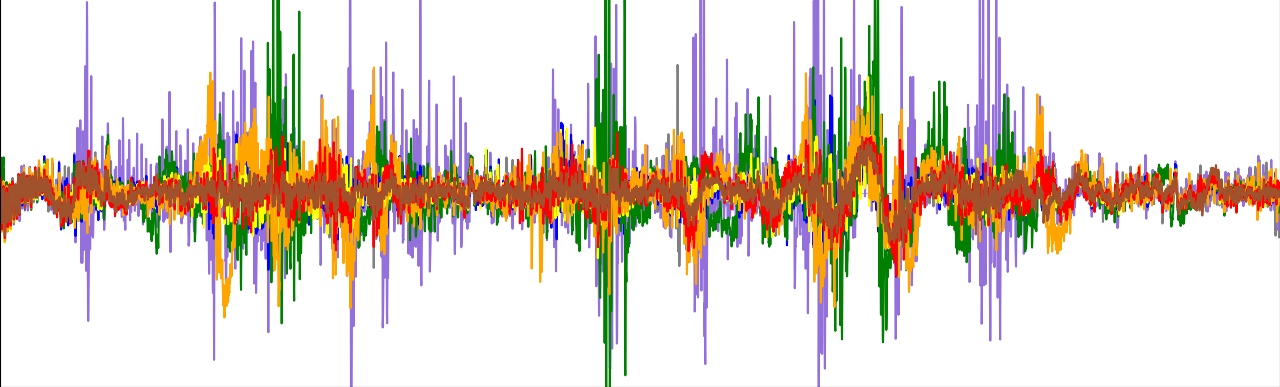}
    \caption{{\large $E_V$ } - EMG from vocalized speech}
    \end{subfigure}
    \begin{subfigure}[b]{0.5\textwidth}
    \includegraphics[width=0.9\columnwidth]{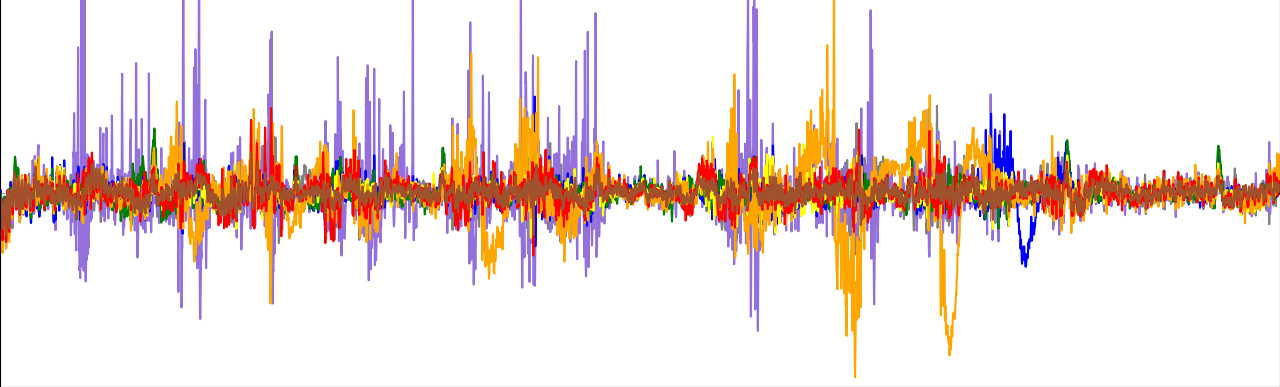}
    \caption{{\large $E_S$ } - EMG from silent speech}
    \end{subfigure}
    
    \caption{The three components of our data that we will use in our model.  The vocalized speech signals $A_V$ and $E_V$ are collected simultaneously and so are time-aligned, while the silent signal $E_S$ is a separate recording of the same utterance without vocalization.  During training we use all three signals, and during testing we are given just $E_S$, from which we must generate audio.  Colors represent different electrodes in the EMG data.  Note that the silent EMG signal $E_S$ is qualitatively different from its vocalized counterpart $E_V$. Not pictured, but also included in our data are the utterance texts, in this case: ``It is possible that the infusoria under the microscope do the same.'' (from H.G. Well's \textit{The War of the Worlds}).}
    \label{fig:example}
\end{figure*}

To capture information about articulator movement, we make use of surface electromyography (EMG).  Surface EMG uses electrodes placed on top of the skin to measure electrical potentials caused by nearby muscle activity.
By placing electrodes around the face and neck, we are able to capture signals from muscles in the speech articulators.  Figure~\ref{fig:emg-pic} shows the EMG electrodes used to capture signals, and Figure~\ref{fig:example} shows an example of EMG signals captured.  We collect EMG measurements during both vocalized speech (normal speech production that has voicing, frication, and other speech sounds) and silent speech (speech-like articulations which do not produce sound).  We denote these EMG signals $E_V$ and $E_S$, respectively.  During the vocalized speech we can also record audio $A_V$, but during silent speech there is no meaningful audio to record.

A substantial body of prior work has explored the use of facial EMG for silent speech-to-text interfaces \cite{Jou2006Towards,schultz2010modeling,kapur2018alterego,meltzner2018development}.  Several initial attempts have also been made to convert EMG signals to speech, similar to the task we approach in this paper \cite{Toth2009Synthesizing,janke2017,diener2018}.  However, these works have focused on the artificial task of recovering audio from EMG that was recorded during \emph{vocalized} speech, rather than the end-goal task of generating from silent speech.  In terms of signals in Figure~\ref{fig:example}, prior work learned a model for producing audio $A_V$ from vocalized EMG $E_V$ and tested primarily on other vocalized EMG signals.  While one might hope that a model trained in this way could directly transfer to silent EMG $E_S$, \citet{Toth2009Synthesizing} show that such a transfer causes a substantial degradation in quality, which we confirm in Section~\ref{sec:experiments}.
This direct transfer from vocalized models fails to account for differences between features of the two speaking modes, such as a lack of voicing in the vocal folds and other changes in articulation to suppress sound.

In this paper, we extend digital voicing to train on silent EMG $E_S$ rather than only vocalized EMG $E_V$.
Training with silent EMG is more challenging than with vocalized EMG, because when training on vocalized EMG data we have both EMG inputs and time-aligned speech targets, but for silent EMG any recorded audio will be silent.
Our solution is to adopt a target-transfer approach, where audio output targets are transferred from vocalized recordings to silent recordings of the same utterances.
We align the EMG features of the instance pairs with dynamic time warping \cite{rabiner1993}, then make refinements to the alignments using canonical correlation analysis \cite{Hotelling1936Relations} and audio feature outputs from a partially trained model.  The alignments can then be used to associate speech outputs with the silent EMG signals $E_S$, and these speech outputs are used as targets for training a recurrent neural transduction model.

We validate our method using both human and automatic metrics, and find that a model trained with our target transfer approach greatly outperforms a model trained on vocalized EMG alone.
On a closed-vocabulary domain (date and time expressions \S\ref{sec:closed-data}), transcription word error rate (WER) from a human evaluation improves from 64\% to just 4\%.  On a more challenging open vocabulary domain (reading from books \S\ref{sec:open-data}) intelligibility measurements improve by 20\% -- from 88\% to 68\% with automatic transcription or 95\% to 75\% with human transcription.

We release our dataset of EMG signals collected during both silent and vocalized speech.  The dataset contains nearly 20 hours of facial EMG signals from a single speaker.  To our knowledge, the largest public EMG-speech dataset previously available contains just 2 hours of data \cite{Wand2014UKA}, and many papers continue to use private datasets.
We hope that this public release will encourage development on the task and allow for fair comparisons between methods.

\section{Data Collection}
\label{sec:data}

We collect a dataset of EMG signals and time-aligned audio from a single speaker during both silent and vocalized speech.
Figure~\ref{fig:example} shows an example from the data collected.
The primary portion of the dataset consists of parallel silent / vocalized data, where the same utterances are recorded using both speaking modes.  These examples can be viewed as tuples $(E_S,E_V,A_V)$ of silent EMG, vocalized EMG, and vocalized audio, where $E_V$ and $A_V$ are time-aligned.  Both speaking modes of an utterance were collected within a single session to ensure that electrode placement is consistent between them.  For some utterances, we record only the vocalized speaking mode.  We refer to these instances as non-parallel data, and represent them with the tuple $(E_V,A_V)$.
Examples are segmented at the utterance level.  The text that was read is included with each instance in the dataset, and is used as a reference when evaluating intelligibility in Section~\ref{sec:experiments}.

For comparison, we record data from two domains: a closed vocabulary and open vocabulary condition, which are described in Sections~\ref{sec:closed-data} and \ref{sec:open-data} below.  Section~\ref{sec:data-details} then provides additional details about the recording setup.

\subsection{Closed Vocabulary Condition}
\label{sec:closed-data}

Like other speech-related signals, the captured EMG signals from a particular phoneme may look different depending on its context.  For this reason, our initial experiments will use a more focused vocabulary set before expanding to a large vocabulary in Section~\ref{sec:open-data} below.

To create a closed-vocabulary data condition, we generate a set of date and time expressions for reading.  These expressions come from a small set of templates such as ``\texttt{<weekday> <month> <year>}'' which are filled in with randomly selected values (over 50,000 unique utterances are possible from this scheme).  Table~\ref{tab:closed-data-summary} summarizes the properties of the data collected in this condition.
A validation set of 30 utterances and a test set of 100 utterances are selected randomly, leaving 370 utterances for training.

\begin{table}
    \centering
    \begin{tabular}{l}
    \toprule
    \textbf{Closed Vocabulary Condition} \\ \midrule
    \enskip \textbf{Parallel silent / vocalized speech} \\
    \quad \quad $(E_S,E_V,A_V)$ \\
    \quad 26 minutes silent / 30 minutes vocalized \\
    \quad Single session \\
    \quad 500 utterances \\
    \quad Average of 4 words per utterance \\
    \quad 67 words in vocabulary \\
    \bottomrule
    \end{tabular}
    \caption{Closed vocabulary data summary}
    \label{tab:closed-data-summary}
\end{table}

\subsection{Open Vocabulary Condition}
\label{sec:open-data}

The majority of our data was collected with open-vocabulary sentences from books.  We use public domain books from Project Gutenberg.\footnote{https://www.gutenberg.org/}
Unlike the closed-vocabulary data which is collected in a single sitting, the open-vocabulary data is broken into multiple sessions where electrodes are reattached before each session and may have minor changes in position between different sessions.
In addition to sessions with parallel silent and vocalized utterances, we also collect non-parallel sessions with only vocalized utterances.
A summary of dataset features is shown in Table~\ref{tab:open-data-summary}.
We select a validation and test set randomly from the silent parallel EMG data, with 30 and 100 utterances respectively.  Note that during testing, we use only the silent EMG recordings $E_S$, so the vocalized recordings of the test utterances are unused.

\begin{table}
    \centering
    \begin{tabular}{l}
    \toprule
    \textbf{Open Vocabulary Condition} \\ \midrule
    \enskip \textbf{Parallel Silent / Vocalized Speech} \\
    \quad \quad $(E_S,E_V,A_V)$ \\
    \quad 3.6 hours silent / 3.9 hours vocalized \\
    \quad Average session has 30 min. of each mode \\
    \quad 1588 utterances \\
    \enskip \textbf{Non-parallel Vocalized Speech} \\
    \quad \quad $(E_V,A_V)$ \\
    \quad 11.2 hours \\
    \quad Average session length 67 minutes \\
    \quad 5477 utterances \\ \midrule
    \enskip \textbf{Total} \\
    \quad 18.6 hours \\
    \quad Average of 16 words per utterance \\
    \quad 9828 words in vocabulary \\
    \bottomrule
    \end{tabular}
    \caption{Open vocabulary data summary}
    \label{tab:open-data-summary}
\end{table}

\subsection{Recording Details}
\label{sec:data-details}

EMG signals are recorded using an OpenBCI Cyton Biosensing Board and transmitted to a computer over WiFi.  Eight channels are collected at a sample rate of 1000 Hz.  The electrode locations are described in Table~\ref{tab:electrode-locations}.  Gold-plated electrodes are used with Ten20 conductive electrode paste.  We use a monopolar electrode configuration, with a shared reference electrode behind one ear.  An electrode connected to the Cyton board's bias pin is placed behind the other ear to actively cancel common-mode interference.  A high pass Butterworth filter with cutoff 2 Hz is used to remove offset and drift in the collected signals, and AC electrical noise is removed with notch filters at 60 Hz and its harmonics.  Forward-backward filters are used to avoid phase delay.

Audio is recorded from a built-in laptop microphone at 16kHz.  Background noise is reduced using a spectral gating algorithm,\footnote{https://pypi.org/project/noisereduce/} and volume is normalized across sessions based on peak root-mean-square levels.

\begin{table}
\centering
\begin{tabular}{ll}
\toprule
& \textbf{Location} \\ \midrule
1 & left cheek just above mouth \\
2 & left corner of chin \\
3 & below chin back 3 cm \\
4 & throat 3 cm left from Adam's apple \\
5 & mid-jaw right \\
6 & right cheek just below mouth \\
7 & right cheek 2 cm from nose \\
8 & back of right cheek, 4 cm in front of ear \\
ref & below left ear \\
bias & below right ear \\
\bottomrule
\end{tabular}
\caption{Electrode locations.}
\label{tab:electrode-locations}
\end{table}

\section{Method}

Our method is built around a recurrent neural transduction model from EMG features to time-aligned speech features (Section~\ref{sec:model-transducer}).
We will denote the featurized version of the signals used by the transduction model $E_{S/V}'$ and $A_V'$ for EMG and audio respectively.
When training solely on vocalized EMG data ($E_V'$ to $A_V'$), training this model is straightforward.
However, our experiments show that training on vocalized EMG alone leads to poor performance when testing on silent EMG (Section~\ref{sec:experiments}) because of differences between the two speaking modes.

A core contribution of our work is a method of training the transducer model on silent EMG signals, which no longer have time-aligned audio to use as training targets.
We briefly describe our method here, then refer to section Section~\ref{sec:model-alignment} for more details.
Using a set of utterances recorded in both silent and vocalized speaking modes, we find alignments between the two recordings and use them to
associate speech features from the vocalized instance ($A_V'$) with the silent EMG $E_S'$.
The alignment is initially found using dynamic time warping between EMG signals and then is refined using canonical correlation analysis (CCA) and predicted audio from a partially trained model.

Finally, to generate audio from predicted speech features, we use a WaveNet decoder, as described in Section~\ref{sec:model-wavenet}.

\subsection{EMG to Speech Feature Transducer}
\label{sec:model-transducer}

When converting EMG input signals to audio outputs, our first step is to use a bidirectional LSTM to convert between featurized versions of the signals, $E'$ and $A'$.
Both feature representations operate at the same frequency, 100 Hz, so that each EMG input $E'[i]$ corresponds to a single time-aligned output $A'[i]$.
Our primary features for representing EMG signals are the time domain features from \citet{Jou2006Towards}, which are commonly used in the EMG-speech-to-text literature.
After splitting the signal from each channel into low and high-frequency components ($x_{low}$ and $x_{high}$) using a triangular filter with cutoff 134 Hz, the signal is windowed with a frame length of 27 ms and shift of 10 ms.  For each frame, five features are computed as follows:
\begin{equation*}
\left[\frac{1}{n}\sum_i (x_{low}[i])^2,\  \frac{1}{n}\sum_i x_{low}[i],\ \frac{1}{n}\sum_i (x_{high}[i])^2, \right.
\end{equation*}
\begin{equation*}
\left. \frac{1}{n}\sum_i |x_{high}[i]|,\ \text{ZCR}(x_{high}) \right]
\end{equation*}
where ZCR is the zero-crossing rate. 
In addition to the time domain features, we also append magnitude values from a 16-point Short-time Fourier transform for each 27 ms frame, which gives us 9 additional features.  The two representations result in a total of 112 features to represent the 8 EMG channels.
Speech is represented with 26 Mel-frequency cepstral coefficients (MFCCs) from 27 ms frames with 10 ms stride.
All EMG and audio features are normalized to approximately zero mean and unit variance before processing.
To help the model to deal with minor differences in electrode placement across sessions, we represent each session with a 32 dimensional session embedding and append the session embedding to the EMG features across all timesteps of an example before feeding into the LSTM.

The LSTM model itself consists of 3 bidirectional LSTM layers with 1024 hidden units, followed by a linear projection to the speech feature dimension.  Dropout $0.5$ is used between all layers, as well as before the first LSTM and after the last LSTM.  The model is trained with a mean squared error loss against time-aligned speech features using the Adam optimizer.
The initial learning rate is set to $.001$, and is decayed by half after every 5 epochs with no improvement in validation loss.
We evaluate a loss on the validation set at the end of every epoch, and select the parameters from the epoch with the best validation loss as the final model.

\subsection{Audio Target Transfer}
\label{sec:model-alignment}

\begin{figure}
    \centering
    \includegraphics[width=.9\columnwidth]{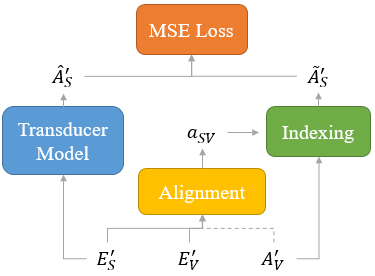}
    \caption{Our audio target transfer method for training on silent EMG $E_S$.  Details in Section~\ref{sec:model-alignment}.}
    \label{fig:model-alignment}
\end{figure}

To train the EMG to speech feature transducer, we need speech features that are time-aligned with the EMG features to use as target outputs.  However, when training with EMG from silent speech, simultaneously-collected audio recordings do not have any audible speech to use as targets.  In this section, we describe how parallel utterances, as described in Section~\ref{sec:data}, can be used to transfer audio feature labels from a vocalized recording to a silent one.  More concretely, given a tuple $(E_S',E_V',A_V')$ of features from silent speech EMG, vocalized speech EMG, and vocalized speech audio, where $E_V$ and $A_V$ are collected simultaneously, we estimate a set of audio features $\tilde{A}_S'$ that time-align with $E_S'$ and represent the output that we would like our transduction network to predict.  A diagram of the method can be found in Figure~\ref{fig:model-alignment}.

Our alignment will make use of dynamic time warping (DTW) \cite{rabiner1993}, a dynamic programming algorithm for finding a minimum-cost monotonic alignment between two sequences $s_1$ and $s_2$.  DTW builds a table $d[i,j]$ of the minimum cost of alignment between the first $i$ items in $s_1$ and the first $j$ items in $s_2$.  The recursive step used to fill this table is $d[i,j]=\delta[i,j]+\min\left(d[i-1,j],d[i,j-1],d[i-1,j-1]\right)$, where $\delta[i,j]$ is the local cost of aligning $s_1[i]$ with $s_2[j]$.
After the dynamic program, we can follow backpointers through the table to find a path of $(i,j)$ pairs representing an alignment.
Although the path is monotonic, a single position $i$ may repeat several times with increasing values of $j$.  We take the first pair from any such sequence to form a mapping $a_{s_1s_2}[i]\rightarrow j$ from every position $i$ in $s_1$ to a position $j$ in $s_2$.

For our audio target transfer, we perform DTW as described above with $s_1=E_S'$ and $s_2=E_V'$.
Initially, we use euclidean distance between the features of $E_S'$ and $E_V'$ for the alignment cost ($\delta_{\text{EMG}}[i,j]=\norm{E_S'[i]-E_V'[j]}$), but will describe several refinements to this choice in Sections~\ref{sec:cca} and \ref{sec:audio-alignment} below.
DTW results in an alignment $a_{SV}[i]\rightarrow j$ that tells us a position $j$ in $E_V'$ for every position $i$ in $E_S'$.  We can then create a warped audio feature sequence $\tilde{A}_S'$ that aligns with $E_S'$ using $\tilde{A}_S'[i]=A_V'[a_{SV}[i]]$.  During training of the EMG to audio transduction model, we will use $\tilde{A}_S'$ as our targets for the transduction outputs $\hat{A}_S'$ when calculating a loss.

This procedure of aligning signals to translate between them is reminiscent of some DTW-based methods for the related task of voice conversion \cite{kobayashi2018sprocket,desai2009voice}.  The difference between these tasks is that our task operates on triples $(E_S,E_V,A_V)$ and must account for the difference in modality between the input $E_S$ and output $A_V$,
while voice conversion operates in a single modality with examples of the form $(A_1,A_2)$.

In addition to training the transducer from $E_S'$ to $\tilde{A}_S'$, we also find that training on the vocalized signals ($E_V'$ to $A_V'$) improves performance.  The vocalized samples are labeled with different session embeddings to allow the model to specialize to each speaking mode.
Each training batch contains samples from both modes mixed together.
For the open vocabulary setting, the full set of examples to sample from has 3 sources: $(E_S',\tilde{A}_S')$ created from parallel utterances, $(E_V,A_V)$ from the vocalized recording of the parallel utterances, and $(E_V,A_V)$ from the non-parallel vocalized recordings.

\subsubsection{CCA}
\label{sec:cca}

While directly aligning EMG features $E_S'$ and $E_V'$ can give us a rough alignment between the signals, doing so ignores the differences between the two signals that lead us to want to train on the silent signals in the first place (e.g. inactivation of the vocal folds and changes in manner of articulation to prevent frication).  To better capture correspondences between the signals, we use canonical correlation analysis (CCA) \cite{Hotelling1936Relations} to find components of the two signals which are more highly correlated.
Given a number of paired vectors $(v_1,v_2)$, CCA finds linear projections $P_1$ and $P_2$ that maximize correlation between corresponding dimensions of $P_1v_1$ and $P_2v_2$.

To get the initial pairings required by CCA, we use alignments found by DTW with the raw EMG feature distance $\delta_\text{EMG}$.  We aggregate aligned $E_S'$ and $E_V'$ features over the entire dataset and feed these to a CCA algorithm to get projections $P_S$ and $P_V$.  CCA allows us to choose the dimensionality of the space we are projecting to, and we use 15 dimensions for all experiments.
Using the projections from CCA, we define a new cost for DTW $$\delta_\text{CCA}[i,j]=\norm{P_SE_S'[i]-P_VE_V'[j]}$$

Our use of CCA for DTW is similar to \citet{zhou2009canonical}, which combined the two methods for use in aligning human pose data, but we found their iterative approach did not improve performance compared to a single application of CCA in our setting.

\subsubsection{Refinement with Predicted Audio}
\label{sec:audio-alignment}

So far, our alignments between the silent and vocalized recordings have relied solely on distances between EMG features.  In this section, we propose an additional alignment distance term that uses audio features.
Although the silent recording has no useful audio signal, once we start to train a transducer model from $E_S'$ to audio features, we can try to align the predicted audio features $\hat{A}_S'$ to vocalized audio features $A_V'$.  Combining with an EMG-based distance, our new cost for DTW becomes
$$\delta_\text{full}[i,j]=\delta_\text{CCA}[i,j]+\lambda\norm{\hat{A}_S'[i]-A_V'[j]}$$
where $\lambda$ is a hyperparameter to control the relative weight of the two terms.  We use $\lambda=10$ for all experiments in this paper.

When training a transducer model using predicted-audio alignment, we perform the first four epochs using only EMG-based alignment costs $\delta_\text{CCA}$.  Then, at the beginning of the fifth epoch, we use the partially-trained transducer model to compute alignments with cost $\delta_\text{full}$.  From then on, we re-compute alignments every five epochs of training.

\subsection{WaveNet Synthesis}
\label{sec:model-wavenet}

To synthesize audio from speech features, we use a WaveNet decoder \cite{Oord2016WaveNet}, which generates the audio sample by sample conditioned on MFCC speech features $A'$.
WaveNet is capable of generating fairly natural sounding speech, in contrast to the vocoder-based synthesizer used in previous EMG-to-speech papers, which caused significant degradation in naturalness \cite{janke2017}.
Our full synthesis model consists of a bidirectional LSTM of 512 dimensions, a linear projection down to 128 dimensions, and finally the WaveNet decoder which generates samples at 16 kHz.  We use a WaveNet implementation from NVIDIA\footnote{https://github.com/NVIDIA/nv-wavenet} which provides efficient GPU inference.
WaveNet hyperparameters can be found in Appendix~\ref{sec:wavenet-hypers}.
During training, the model is given gold speech features as input, which we found to work better than training from EMG-predicted features.
Due to memory constraints we do not use any batching during training, but other optimization hyperparameters are the same as those from Section~\ref{sec:model-transducer}.

\section{Experiments}
\label{sec:experiments}

In this section, we run experiments to measure intelligibility of audio generated by our model from silent EMG signals $E_S$.  Since prior work has trained only on vocalized EMG signals $E_V$, we compare our method to a \emph{direct transfer} baseline which trains a transducer model only on vocalized EMG $E_V$ before testing on the silent EMG $E_S$.\footnote{Note that because prior work has not released data or code, we are unable to perform a direct comparison to experiments found in their papers.  Our direct transfer baseline represents a conceptually equivalent model, but with larger neural networks than prior work.}
The baseline transducer and wavenet models have identical architecture to those used by our method, but are not trained with silent EMG using our target transfer approach.
Since one may hypothesize that most of the differences between silent and vocalized EMG will take place near the vocal folds, we also test a variant of this baseline where the electrode placed on the neck is ignored.

We first test on the closed vocabulary data described in Section~\ref{sec:closed-data}, then on the open vocabulary data from Section~\ref{sec:open-data}.  On the open vocabulary data, we also run ablations to evaluate different alignment refinements with CCA and predicted audio (see Sections~\ref{sec:cca} and \ref{sec:audio-alignment}).

\subsection{Closed Vocabulary Condition}
\label{sec:closed-data-results}

We begin by testing intelligibility on the closed vocabulary date and time data with a human transcription evaluation.  The human evaluator is given a set of 20 audio output files from each model being tested (listed below) and is asked to write out in words what they heard.
The files to transcribe are randomly shuffled, and the evaluator is not told that the outputs come from different systems.
They are told that the examples will contain dates and times, but are not given any further information about what types of expressions may occur.  The full text of the instructions provided to the evaluator can be found in Appendix~\ref{sec:human-eval-instructions}.
We compare the transcriptions from the human evaluator to the original text prompts that were read during data collection to compute a transcription word error rate (WER):
$$\text{WER}=\frac{\text{substitutions}+\text{insertions}+\text{deletions}}{\text{reference length}}$$
Lower WER values indicate better models.

Using this evaluation, we compare three different models: a direct transfer baseline trained only on vocalized EMG signals, a variant of this baseline where the throat electrode is removed to reduce divergence between speaking modes, and our full model trained on silent EMG using target-transfer.
All three models were trained on open vocabulary data (Section~\ref{sec:open-data}) before being fine-tuned on the closed vocabulary training set.
A single WaveNet model is used to synthesize audio for all three models and was also trained on the open vocabulary data before being fine-tuned in-domain.

The results of our evaluation are shown in Table~\ref{tab:closed-data-results}.
We first observe that removing the throat electrode substantially improves intelligibility for the direct transfer baseline.  Although this modification removes potentially useful information, it also removes divergence between the silent and vocalized EMG signals.  Its relative success further motivates the need for methods to account for the differences in the two modes, such as our target-transfer approach.  However, even with the throat-removal modification, the direct transfer approach is still only partially intelligible.

A model trained with our full approach, including CCA and predicted-audio alignment, achieves a WER of 3.6\%.  This result represents a high level of intelligibility and a 94\% relative error reduction from the strongest baseline.

\begin{table}
    \centering
    \begin{tabular}{lr}
        \toprule
        \textbf{Model} & \textbf{WER} \\ \midrule
        Direct transfer baseline & 88.8 \\
        \quad Without throat electrode & 64.6 \\ \midrule
        Our model & \textbf{3.6} \\ \bottomrule
    \end{tabular}
    \caption{Results of a human intelligibility evaluation on the closed vocabulary data.  Lower WER is better.  Our model greatly outperforms both variants of the direct transfer baseline.}
    \label{tab:closed-data-results}
\end{table}

\subsection{Open Vocabulary Condition}
\label{sec:open-data-results}

Similar to our evaluation in Section~\ref{sec:closed-data-results}, we use a transcription WER to evaluate intelligibility of model outputs in the open vocabulary condition.  For the open vocabulary setting, we evaluate both with a human transcription and with transcriptions from an automatic speech recognizer.

\subsubsection{Human Evaluation}

Our human evaluation with open vocabulary outputs follows the same setup as the closed vocabulary evaluation.  Transcripts are collected for 20 audio outputs from each system, with a random interleaving of outputs from the different systems.  The annotator had no prior information on the content of the texts being evaluated.  We compare two systems: direct transfer without the throat electrode (the stronger baseline) and our full model.

The results of this evaluation are a 95.1\% WER for the direct transfer baseline and 74.8\% WER for our system.  While the intelligibility is much lower than in the closed vocabulary condition, our method still strongly out-performs the baseline with a 20\% absolute improvement.

\subsubsection{Automatic Evaluation}

In addition to the human evaluation, we also perform an automatic evaluation by transcribing system outputs with a large-vocabulary automatic speech recognition (ASR) system.
Using an automatic transcription allows for much faster and more reproducible comparisons between methods compared to a human evaluation.
For our automatic speech recognizer, we use the open source implementation of DeepSpeech from Mozilla\footnote{https://github.com/mozilla/DeepSpeech} \cite{hannun2014deep}.  Running the recognizer on the original vocalized audio recordings from the test set results in a WER of 9.5\%, which represents a lower bound for this evaluation.

Our automatic evaluation results are shown in Table~\ref{tab:open-data-results}.
While the absolute WER values for the ASR evaluation do not perfectly match those of the human transcriptions, both evaluations show a 20\% improvement of our system over the best baseline. Given this correlation between evaluations and the many advantages of automated evaluation, we will use the automatic metric throughout the rest of this work and recommend its use for comparisons in future work.

We also run ablations of the two alignment refinement methods from Sections~\ref{sec:cca} and \ref{sec:audio-alignment} and include results in Table~\ref{tab:open-data-results}.  We see that both refinements have a positive effect on performance, though the impact of aligning with predicted audio is greater.

\begin{table}
    \centering
    \begin{tabular}{lr}
        \toprule
        \textbf{Model} & \textbf{WER} \\ \midrule
        Direct transfer baseline & 91.2 \\
        \quad Without throat electrode & 88.0 \\
        \midrule
        Our model & \textbf{68.0} \\
        \quad Without CCA & 69.8 \\
        \quad Without audio alignment & 76.5 \\
        \bottomrule
    \end{tabular}
    \caption{Results of an automatic intelligibility evaluation on open vocabulary data.  Lower WER is better.}
    \label{tab:open-data-results}
\end{table}

\subsection{Additional Experiments}

In the following subsections, we perform additional experiments on the open vocabulary data to explore the effect of data size and choice of electrode positions.  These experiments are all evaluated using the automatic transcription method described in Section~\ref{sec:open-data-results}.

\subsubsection{Data Size}

In this section we explore the effect of dataset size on model performance.  We train the EMG-to-speech transducer model on various-sized fractions of the dataset, from 10\% to 100\%, and plot the resulting WER.  We select from the parallel (silent and vocalized) and non-parallel (vocalized only) portions proportionally here, but will re-visit the difference later.
Although data size also affects WaveNet quality, we use a single WaveNet trained on the full dataset for all evaluations to focus on EMG-specific data needs.

\begin{figure}
    \centering
    \begin{tikzpicture}
    	\begin{axis}[
    		height=.65\columnwidth,
    		width=.95\columnwidth,
    		xmin=0,
    		xmax=20,
    		ymin=60,
    		ymax=90,
    		xlabel=Data Amount (Hours),
    		ylabel=WER,
    		label style = {font=\small},
    		ticklabel style = {font=\tiny},
    		axis lines* = left,
    		xlabel near ticks,
    		ylabel near ticks
    	]
    
    	\addplot[only marks, color=cornflowerblue] coordinates {
    		(1.86, 87.6)
    		(3.72, 83.2)
    		(5.58, 78.7)
    		(7.44, 76)
    		(9.3, 73.9)
    		(11.16, 73.3)
    		(13.02, 73.1)
    		(14.88, 72)
    		(16.74, 69.3)
    		(18.6, 68)
    	};
    	\end{axis}
    \end{tikzpicture}
    \caption{Effect of data amount on intelligibility.}
    \label{fig:data-size}
\end{figure}
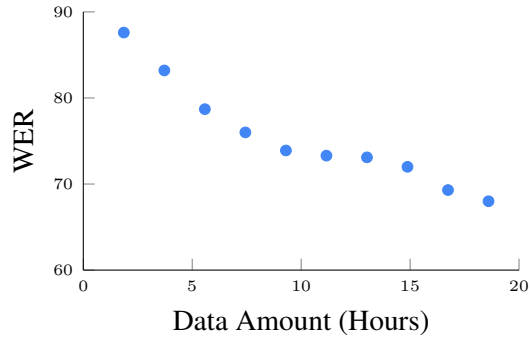

Figure~\ref{fig:data-size} shows the resulting intelligibility measurements for each data size.  As would be expected, the rate of improvement is larger when data sizes are small.  However, there does not seem to be a plateau in performance, as improvements continue even when increasing data size beyond fifteen hours.  These continued gains suggest that collecting additional data could provide more improvement in the future.

We also train a model without the non-parallel vocalized data (vocalized recordings with no associated silent recording; see Section~\ref{sec:data}).  A model trained without this data has a WER of 71.6\%, a loss of 3.6 absolute percentage points.  This confirms that non-parallel vocalized data can be useful for silent speech even though it contains only data from the vocalized speaking mode.  However, if we compare this accuracy to a model where the same amount of data was removed proportionally from the two data types (parallel and non-parallel), we see that removing a mixture of both types leads to a much larger performance decrease to 76\% WER.
This indicates that the non-parallel data is less important to the performance of our model, and suggests that future data collection efforts should focus on collecting parallel utterances of silent and vocalized speech rather than non-parallel utterances of vocalized speech.

\subsubsection{Removing Electrodes}

In this section, we experiment with models that operate on a reduced set of electrodes
to assess the impact on performance and gain information about which electrodes are most important.  We perform a random search to try to find a subset of four electrodes that works well.  More specifically, we sample 10 random combinations of four electrodes to remove (out of 70 possible combinations) and train a model with each.  We then use validation loss to select the best models.

The three best-performing models removed the following sets of electrodes (using electrode numbering from Table~\ref{tab:electrode-locations}): 1) $\{4,5,7,8\}$ 2) $\{3,5,7,8\}$ and 3) $\{2,5,7,8\}$.  We note that electrodes 5, 7, and 8 (which correspond with electrodes on the mid-jaw, upper cheek, and back cheek) appear in all of these, indicating that they may be contributing less to the performance of the model.
However, the best model we tested with four electrodes did have substantially worse intelligibility compared to an eight-electrode model, with 76.8\% WER compared to 68.0\%. A model that removed only electrodes 5, 7, and 8 also performed substantially worse, with a WER of 75.3\%.

\section{Conclusion}

Our results show that digital voicing of silent speech, while still challenging in open domain settings, shows promise as an achievable technology.
We show that it is important to account for differences in EMG signals between silent and vocalized speaking modes and demonstrate an effective method of doing so.
On silent EMG recordings from closed vocabulary data our speech outputs achieve high intelligibility, with a 3.6\% transcription word error rate and relative error reduction of 95\% from our baseline.  We also significantly improve intelligibility in an open vocabulary condition, with a relative error reduction over 20\%.  We hope that our public release of data will encourage others to further improve models for this task.\footnote{Our dataset can be downloaded from https://doi.org/10.5281/zenodo.4064408 and code is available at https://github.com/dgaddy/silent\_speech.}

\section*{Acknowledgments}

This material is based upon work supported by the National Science Foundation under Grant No. 1618460.

\bibliographystyle{acl_natbib}
\bibliography{references}

\begin{thebibliography}{16}
\expandafter\ifx\csname natexlab\endcsname\relax\def\natexlab#1{#1}\fi

\bibitem[{Desai et~al.(2009)Desai, Raghavendra, Yegnanarayana, Black, and
  Prahallad}]{desai2009voice}
Srinivas Desai, E~Veera Raghavendra, B~Yegnanarayana, Alan~W Black, and Kishore
  Prahallad. 2009.
\newblock Voice conversion using artificial neural networks.
\newblock In \emph{2009 IEEE International Conference on Acoustics, Speech and
  Signal Processing}, pages 3893--3896. IEEE.

\bibitem[{{Diener} et~al.(2018){Diener}, {Felsch}, {Angrick}, and
  {Schultz}}]{diener2018}
L.~{Diener}, G.~{Felsch}, M.~{Angrick}, and T.~{Schultz}. 2018.
\newblock Session-independent array-based {EMG}-to-speech conversion using
  convolutional neural networks.
\newblock In \emph{Speech Communication; 13th ITG-Symposium}, pages 1--5.

\bibitem[{Hannun et~al.(2014)Hannun, Case, Casper, Catanzaro, Diamos, Elsen,
  Prenger, Satheesh, Sengupta, Coates et~al.}]{hannun2014deep}
Awni Hannun, Carl Case, Jared Casper, Bryan Catanzaro, Greg Diamos, Erich
  Elsen, Ryan Prenger, Sanjeev Satheesh, Shubho Sengupta, Adam Coates, et~al.
  2014.
\newblock Deep speech: Scaling up end-to-end speech recognition.
\newblock \emph{arXiv preprint arXiv:1412.5567}.

\bibitem[{Hotelling(1936)}]{Hotelling1936Relations}
Harold Hotelling. 1936.
\newblock Relations between two sets of variates.

\bibitem[{{Janke} and {Diener}(2017)}]{janke2017}
M.~{Janke} and L.~{Diener}. 2017.
\newblock {EMG}-to-speech: Direct generation of speech from facial
  electromyographic signals.
\newblock \emph{IEEE/ACM Transactions on Audio, Speech, and Language
  Processing}, 25(12):2375--2385.

\bibitem[{Jou et~al.(2006)Jou, Schultz, Walliczek, Kraft, and
  Waibel}]{Jou2006Towards}
Szu-Chen~Stan Jou, Tanja Schultz, Matthias Walliczek, Florian Kraft, and
  Alexander~H. Waibel. 2006.
\newblock Towards continuous speech recognition using surface electromyography.
\newblock In \emph{INTERSPEECH}.

\bibitem[{Kapur et~al.(2018)Kapur, Kapur, and Maes}]{kapur2018alterego}
Arnav Kapur, Shreyas Kapur, and Pattie Maes. 2018.
\newblock Alterego: A personalized wearable silent speech interface.
\newblock In \emph{23rd International Conference on Intelligent User
  Interfaces}, pages 43--53.

\bibitem[{Kobayashi and Toda(2018)}]{kobayashi2018sprocket}
Kazuhiro Kobayashi and Tomoki Toda. 2018.
\newblock sprocket: Open-source voice conversion software.
\newblock In \emph{Odyssey}, pages 203--210.

\bibitem[{Meltzner et~al.(2017)Meltzner, Heaton, Deng, De~Luca, Roy, and
  Kline}]{meltzner2017silent}
Geoffrey~S Meltzner, James~T Heaton, Yunbin Deng, Gianluca De~Luca, Serge~H
  Roy, and Joshua~C Kline. 2017.
\newblock Silent speech recognition as an alternative communication device for
  persons with laryngectomy.
\newblock \emph{IEEE/ACM transactions on audio, speech, and language
  processing}, 25(12):2386--2398.

\bibitem[{Meltzner et~al.(2018)Meltzner, Heaton, Deng, De~Luca, Roy, and
  Kline}]{meltzner2018development}
Geoffrey~S Meltzner, James~T Heaton, Yunbin Deng, Gianluca De~Luca, Serge~H
  Roy, and Joshua~C Kline. 2018.
\newblock Development of {sEMG} sensors and algorithms for silent speech
  recognition.
\newblock \emph{Journal of neural engineering}, 15(4):046031.

\bibitem[{van~den Oord et~al.(2016)van~den Oord, Dieleman, Zen, Simonyan,
  Vinyals, Graves, Kalchbrenner, Senior, and Kavukcuoglu}]{Oord2016WaveNet}
A{\"a}ron van~den Oord, Sander Dieleman, Heiga Zen, Karen Simonyan, Oriol
  Vinyals, Alex Graves, Nal Kalchbrenner, Andrew~W. Senior, and Koray
  Kavukcuoglu. 2016.
\newblock Wave{N}et: A generative model for raw audio.
\newblock \emph{ArXiv}, abs/1609.03499.

\bibitem[{Rabiner and Juang(1993)}]{rabiner1993}
Lawrence Rabiner and Biing-Hwang Juang. 1993.
\newblock \emph{Fundamentals of speech recognition}.
\newblock Prentice Hall.

\bibitem[{Schultz and Wand(2010)}]{schultz2010modeling}
Tanja Schultz and Michael Wand. 2010.
\newblock Modeling coarticulation in {EMG}-based continuous speech recognition.
\newblock \emph{Speech Communication}, 52(4):341--353.

\bibitem[{Toth et~al.(2009)Toth, Wand, and Schultz}]{Toth2009Synthesizing}
Arthur~R. Toth, Michael Wand, and Tanja Schultz. 2009.
\newblock Synthesizing speech from electromyography using voice transformation
  techniques.
\newblock In \emph{INTERSPEECH}.

\bibitem[{Wand et~al.(2014)Wand, Janke, and Schultz}]{Wand2014UKA}
Michael Wand, Matthias Janke, and Tanja Schultz. 2014.
\newblock The {EMG-UKA} corpus for electromyographic speech processing.
\newblock In \emph{INTERSPEECH}.

\bibitem[{Zhou and Torre(2009)}]{zhou2009canonical}
Feng Zhou and Fernando Torre. 2009.
\newblock Canonical time warping for alignment of human behavior.
\newblock In \emph{Advances in neural information processing systems}, pages
  2286--2294.

\end{thebibliography}

\appendix

\section{WaveNet Hyperparameters}
\label{sec:wavenet-hypers}

{
    \centering
    \begin{tabular}{lr}
        Hyperparameter & Value \\ \hline
        n\_in\_channels & 256 \\
        n\_layers & 16 \\
        max\_dilation & 128 \\
        n\_residual\_channels & 64 \\
        n\_skip\_channels & 256 \\
        n\_out\_channels & 256 \\
        n\_cond\_channels & 128 \\
        upsamp\_window & 432 \\
        upsamp\_stride & 160 
    \end{tabular}
}

\section{Human Evaluator Instructions}
\label{sec:human-eval-instructions}

The instructions given to the human evaluator are as follows:
``Please listen to each of the attached sound files and write down what you hear as best you can.  There are 60 files, each of which will contain an expression of some date or time.  Write your transcriptions into a spreadsheet such as Excel or Google sheets so that the row numbers match the numbers in the file names.  Although many of the clips will contain numbers, please write out what you hear as words.  For example, you might write something like:
\texttt{five oh two pm on Thursday}\footnote{We intentionally used an example that does not match a pattern in our generation procedure to avoid biasing the evaluator.}
Many of the clips may be difficult to hear.  If this is the case, write whatever words you are able to make out, even if it does not form a complete expression.  For example:
\texttt{five two pm on}
If you cannot make out any words, leave the corresponding row blank.''

\section{Additional Data Collection Details}

During data collection, text prompts consisting of a single sentence to be read are displayed on a screen.  After reading the sentence, the subject pressed a key to advance to the next sentence.  If they were unhappy with a recording, they could press another key to re-record an utterance.  A real-time display of EMG signals was used to monitor the electrodes for excessive noise.  During silent speech, the subject was instructed to mouth words as naturally as possible without producing sound.

\section{Additional Reproducibility Information}

Models were trained for up to two days on a single K80 GPU.
Hyperparameter search consisted of a mixture of manual and random search, typically with less than 10 runs.
Hyperparameters were chosen primarily based validation loss, with major design decisions also being checked with automatic transcription evaluation.

\end{document}